\documentclass[12pt]{iopart}
\usepackage{graphicx}

\begin{document}

\paper[Hidden symmetries and equilibrium\ldots]{ Hidden symmetries and
equilibrium properties 
of multiplicative white--noise stochastic processes}

\author{Zochil Gonz\'alez Arenas$^{1,2}$ and Daniel G.\ Barci$^2$}

\address{$^1$ Instituto de Cibern\'etica, Matem\'atica y F\'\i sica (ICIMAF)\\
Calle 15 \# 551  e/ C y D, Vedado, C. Habana, Cuba.}

\address{$^2$ Departamento de F{\'\i}sica Te\'orica,
Universidade do Estado do Rio de Janeiro, Rua S\~ao Francisco Xavier 524, 20550-013,  Rio de Janeiro, RJ, Brazil.}

\date{\today}

\ead{daniel.barci@gmail.com}

\begin{abstract}
Multiplicative white-noise stochastic processes continuously attract the attention of a wide area of scientific 
research. The variety of prescriptions available to define it difficults the development of general tools for its
characterization. In this work, we study equilibrium properties of Markovian multiplicative white-noise processes.
For this, we define the time reversal transformation for this kind of processes, taking into account that the asymptotic
stationary probability distribution depends on the prescription.
Representing the stochastic process in a functional Grassman formalism, we avoid the necessity of fixing a 
particular prescription. In this framework, we analyze equilibrium properties
and study hidden symmetries of the 
process. We show that, using a careful definition of equilibrium distribution and 
taken into account the appropriate  time reversal transformation, usual equilibrium properties are satisfied 
for any prescription. Finally, we present a detailed deduction of a covariant supersymmetric formulation of a
multiplicative Markovian white-noise process and study some of  the constraints it imposes on correlation 
functions using Ward-Takahashi identities.
\end{abstract}

%02.50.-r 	Probability theory, stochastic processes, and statistics (see also section 05 Statistical physics, thermodynamics, and nonlinear dynamical systems)
 
%02.50.Cw 	Probability theory
 
%02.50.Ey 	Stochastic processes
 
%02.50.Fz 	Stochastic analysis
 
%02.50.Ga 	Markov processes 

%05.10.-a 	Computational methods in statistical physics and nonlinear dynamics (see also 02.70.-c in mathematical methods in physics)
 
%05.10.Cc 	Renormalization group methods
 
%05.10.Gg 	Stochastic analysis methods (Fokker-Planck, Langevin, etc.) 

%05.40.-a 	Fluctuation phenomena, random processes, noise, and Brownian motion (for fluctuations in superconductivity, see 74.40.-n; for statistical theory and fluctuations in nuclear reactions, see 24.60.-k; for fluctuations in plasma, see 52.25.Gj; for nonlinear dynamics and chaos, see 05.45.-a)
 
%05.40.Ca 	Noise 

% 60H40 White noise theory 
% 82B30 Statistical thermodynamics (See also 80-XX) 
% 82B31 Stochastic methods 
% 82C31 Stochastic methods (Fokker-Planck, Langevin, etc.) (See also 60H10) 

\pacs{05.40.-a, 02.50.Ey, 05.10.Gg}

\ams{60H40,82B30, 82B31, 82C31}

\maketitle

%\tableofcontents

%%%%%%%%%%%%%%%%%%%%%%%%%%%%%%%%%%%%%%%%%%%%%%%%%%%%%%
\section{Introduction}
%%%%%%%%%%%%%%%%%%%%%%%%%%%%%%%%%%%%%%%%%%%%%%%%%%%%%%

From the early studies of Brownian motion, more than a hundred years
ago~\cite{Hanggi100years}, 
the use of stochastic differential equations to model a wide variety
of dynamical systems  has grown dramatically. 
Applications can be found along a wide area of
scientific research, from physics and chemistry~\cite{vanKampen,Gardiner},
through biology and eco\-lo\-gy~\cite{Poschel,Murray}, to economy and social
sciences~\cite{Mantegna,Bouchaud}.

In the original formulation of Brownian motion, the
influence of the medium on a diffusive particle is modeled by splitting its
effect in two parts: a deterministic one, given by an homogeneous viscous
force, and a
stochastic part, given by a random force with zero expectation value. In this
way, fluctuations  exhibit as an {\em additive} noise and,
consequently, the considered model represents  an {\em additive stochastic
process} described by a Langevin equation.  
However, the viscous force could have non-homogeneous
contributions, for instance, in the presence of boundary conditions, such as
a diffusion  of a Brownian particle near a wall~\cite{Lancon2001,Lancon2002}.
If the diffusion is not homogeneous, fluctuations could depend on the
state of the system 
and they can be regarded as the product of a random force and a function of the
state variable. In that situation, {\em multiplicative} noise is defined and the
model is known as a {\em multiplicative stochastic process}. Another
interesting example of multiplicative noise is the stochastic
Landau-Lifshitz-Gilbert equation~\cite{Palacios1998},
used to describe dynamics of classical magnetic moments of individual magnetic
nano-particles. In this case, the noisy fluctuations of the magnetic field
couple the magnetic moment in a multiplicative way.  The classification of 
additive or multiplicative noise should be considered in the more general
context of external and internal noise~\cite{vanKampen,Gitterman}. In the former
case,  dissipation and noise can be considered as two effects with
different microscopic origin. In other words, in the absence of noise, the
classical deterministic system is perfectly well defined. However, in the latter
case, dissipation and fluctuation have the same intrinsic origin and it is not
possible to ``turn off'' one of these effects. 

The theory of stochastic evolution provides a beautiful connection 
between dynamics and statistical physics. In general, Einstein
relation or, more generally, fluctuation-dissipation relations associate
dynamical properties of the system with thermodynamical equilibrium.
However, stochastic dynamics not necessarily model  physical systems, where
the long time evolution should conduce to thermodynamical equilibrium. In fact,
it is possible to have more general stationary state distributions that
represent an equilibrium state in the stochastic dynamical sense, not
related to thermodynamics.
Moreover, stochastic dynamics provide an interesting approach to
out-of-equilibrium statistical mechanics~\cite{crooks1999,seifert2005}.
Interestingly, it is possible to attribute thermodynamical concepts like heat,
entropy or free energy to each trajectory of an stochastic evolution, given rise
to the research field usually called  stochastic
thermodynamics~\cite{seifert2008}. 

In this work, we would like to present a study of  equilibrium properties of
 Markovian multiplicative white-noise processes. A  prototype of these processes is
represented by the Langevin equation~(\ref{eq.Langevin}). It is well known
that, due to the Gaussian white-noise distribution, the continuum limit of
the discretized time evolution of multiplicative processes is not unique. In fact, there are different
prescriptions to perform this limit. Perhaps, the most popular prescriptions
are the It\^o~\cite{Ito} and the Stratonovich~\cite{Stratonovich} ones, each one
producing a different stochastic evolution and forcing different rules of
calculus. We will discuss in detail equilibrium properties in the more general
prescription called Generalized Stratonovich convention~\cite{Hanggi1978} (also
called ``$\alpha$-convention''~\cite{Janssen-RG} in the field theory
literature). In this prescription, a continuum   parameter
$0\le\alpha\le 1$ is defined  in such a way that each value
of the parameter corresponds to a different discretization rule of the
stochastic differential equation. The value $\alpha=0$ corresponds with the
It\^o prescription while $\alpha=1/2$ is the Stratonovich one. 

The concept of equilibrium is tightly related with the concept of
time reversal symmetry. For  this reason, it is essential to carefully define
the time reversal transformation of the stochastic process. 
In multiplicative white-noise processes, the forward and backward stochastic
trajectories evolve with different prescriptions (except when considering Stratonovich convention). 
For instance, if It\^o
convention ($\alpha=0$) is defined for the forward evolution, the
backward trajectory evolves with the so called
H\"anggi-Klimontovich~\cite{Hanggi1978, Hanggi1980, Hanggi1982, Klimontovich}
prescription ($\alpha=1$). In general, we will show that the prescription
$\alpha$ is the time reversal conjugate of $(1-\alpha)$. In addition, since  the
asymptotic stationary probability distribution
also depends on the chosen prescription, the correct definition of
time reversal transformation, compatible with a unique equilibrium
distribution, is quite involved.

It is very useful to use, instead of the Langevin or Fokker-Planck approach, 
a path integral formalism~\cite{arenas2010}, in which the central object of the
theory is a set of  stochastic trajectories.
Using this formalism, we will analyze equilibrium properties,
such as detailed balance relations, microscopic reversibility and entropy
production. We will show that, due to  a careful definition of equilibrium
distribution and taken into account the appropriate  time
reversal transformation, usual equilibrium properties are satisfied for {\em any
value of $\alpha$}. 
Even thought the path integral representation of the stochastic process is
suitable for the computation of correlation functions and responses, 
specific calculations are  very
cumbersome, since each value of $\alpha$ defines different  differentiation  
and integration rules. In particular, the ``chain rule'' to compute
total derivatives of a stochastic process depends on $\alpha$. We make a
generalization of the It\^o formula for any value of the discretization
parameter. In this work, we explicitly show the importance of 
this ``generalized chain rule'',  in the path integral formalism. 
  
The stochastic process can also be represented as a path integral in an
extended functional space,  by introducing auxiliary commutative as well as
anti-commutative Grassman variables. The main advantage of this extension for
multiplicative noisy systems resides in the prescription
independent character of the formulation. While in the original path integral formulation the prescription
appears as a continuum limit ambiguity, in
the functional Grassman formalism, the ambiguity appears in the definition of
equal-time Grassman Green functions~\cite{arenas2010}.   
In this way, provided we do not integrate the Grassman variables,  we can
perform any calculation without specifying a particular  prescription. 
In some sense, all the difficulty introduced by the generalized It\^o calculus
is being codified in a simpler Grassman algebra.

As a by-product,  it is possible to study hidden symmetries of the stochastic process.
Indeed, the class of systems  we have studied in this paper is invariant under
linear
transformations in the  extended functional space. This hidden symmetry, called
supersymmetry (SUSY), was recognized in other  stochastic processes several years
ago~\cite{ParisiSourlas1979,Tsvelik1982,ParisiSourlas1982}. 
SUSY properties have been extensively studied for
additive stochastic processes~\cite{Zinn-Justin,Bouchaud1996} and  for
non-Markovian multiplicative processes~\cite{AronLeticia2010}. 
In another hand, for  Markovian multiplicative white-noise 
systems, important progresses have recently been reported~\cite{arenas2012}. The main difficulty in
the SUSY formulation of a Markov multiplicative process is related with the
great variety of prescriptions available to define the Wiener integral, which produces
several stochastic evolutions with different final steady states. Moreover, time
reversal transformations mix different prescriptions.  

In this paper, we deduce a covariant supersymmetric formulation of a
multiplicative Markov process showing in detail the importance of the
prescription dependent equilibrium distribution and the correct definition of
time reversal transformation. 

Physically, SUSY  encodes equilibrium properties of the system. Some of  the
constraints it imposes on correlation 
functions (Ward-Takahashi identities) are related to fluctuation-dissipation
theorems~\cite{Chaturvedi1984}. This property has acquired a refreshing interest
due to the growing importance of  stochastic out-of-equilibrium 
systems~\cite{Corberi2007}. In this sense, it is possible to understand out-of-equilibrium dynamics as a 
symmetry breaking mechanism. We analyze in detail Ward-Takahashi
identities specially related with  fluctuation-dissipation theorems. We 
show that this theorem is fulfilled independently of the chosen prescription
 to define the process. However, the linear response, fluctuations,  as
well as the equilibrium distribution, do depend on the specific prescription.

The paper is organized as follows: In Sect.~\ref{model} we define our
model by means of a Langevin equation with multiplicative noise. In 
Sect.~\ref{TimeReversal} we carefully study the problem of time reversal
symmetry and the equilibrium probability distribution. With these results, we 
analyze properties such as detailed balance and entropy production
in Sect.~\ref{DetailedBalance}. In Sect.~\ref{SUSY},  we present a covariant
supersymmetric formulation of the stochastic process and we deduce the
fluctuation-dissipation theorem for arbitrary prescriptions.  
Finally, we discuss our conclusions in Sect.~\ref{discussion}, leading for
\ref{appendix} some details of the calculations. 

%%%%%%%%%%%%%%%%%%%%%%%%%%%%%%%%%%%%%%%%%%%%%%%%%%%%%%%%%%%%%%%%%%%%
\section{Multiplicative white--noise stochastic evolution}
\label{model}
%%%%%%%%%%%%%%%%%%%%%%%%%%%%%%%%%%%%%%%%%%%%%%%%%%%%%%%%%%%%%%%%%%%%
The main purpose of this section is to  define the model and to establish the concepts and the notation we use in the rest of the paper. 
For simplicity, we consider a single random variable $x(t)$ satisfying a first order differential equation given by
\begin{equation}
\frac{dx(t)}{dt} = f(x(t)) + g(x(t))\zeta(t),
\label{eq.Langevin} 
\end{equation}
where $\zeta(t)$ is a Gaussian white noise,
\begin{equation}
 \left\langle \zeta(t)\right\rangle   = 0 \;\;\mbox{,}\;\;\;  \left\langle \zeta(t)\zeta(t')\right\rangle = \delta(t-t').
\label{whitenoise}
\end{equation}
The drift term $f(x)$ and  the square root of the diffusion function $g(x)$ are,
 in principle,  arbitrary smooth functions of $x(t)$. 
The only restrictive condition is that $g(x)$ should be ``invertible'', {\em i.
e.},   $g(x)\neq 0, \forall x$. 

As it is very well known, to completely define equation~(\ref{eq.Langevin}), it is necessary to give sense to the ill-defined product $g(x(t))\zeta(t)$, since $\zeta(t)$ is delta correlated. 
The problem can be easily understood looking at the integral
\begin{equation}
\int    g(x(t))\; \zeta(t) dt= \int  g(x(t))\;  dW(t) \ ,
\end{equation} 
where we have defined the Wiener process $W(t)$ as $\zeta(t)=dW(t)/dt$. 
By definition, the Riemann-Stieltjes integral is
\begin{equation}
\int   g(x(t))\;  dW(t)=\lim _{n\to\infty} \sum_{j=1}^n  g(x(\tau_j))(W(t_{j+1})-W(t_j))
\label{eq.Wiener}
\end{equation}
where $\tau_j$ is taken in the interval $[t_j,t_{j+1}]$ and the limit is taken
in the sense of {\em mean-square limit}~\cite{Gardiner}. For a smooth measure
$W(t)$, the limit converges to a unique value, regardless the value of
$\tau_j$. 
However, $W(t)$ is not smooth,  in fact, it is nowhere integrable. In any interval, white noise fluctuates an infinite number of times with infinite variance. 
Therefore, the value of the integral depends on the prescription for the choice of $\tau_j$.
There are several prescriptions to define this integral that can be summarized
in the so called ``generalized Stratonovich prescription''~\cite{Hanggi1978} 
or ``$\alpha$-prescription''~\cite{Janssen-RG}, for which we choose 
\begin{equation}
g(x(\tau_j))=g((1-\alpha)x(t_j)+\alpha x(t_{j+1}))\mbox{~~ with~~} 0\le \alpha \le 1. 
\label{eq.prescription}
\end{equation}
In this way, $\alpha=0$ corresponds with the pre-point It\^o interpretation and $\alpha=1/2$ coincides with the (mid-point) Stratonovich one.  
Moreover, the post-point prescription, $\alpha=1$, is also known as the kinetic
or the H\"anggi-Klimontovich
interpretation~\cite{Hanggi1978, Hanggi1980, Hanggi1982, Klimontovich}.

In principle, each particular choice of $\alpha$ fixes a different stochastic evolution.  
In many physical applications, a weakly colored Gaussian-Markov noise with a finite variance~\cite{Hanggi-shot} is considered. 
In this case, there is no problem with the interpretation of equation~(\ref{eq.Langevin}) and we can take the limit of infinite variance at the end of the calculations. 
This regularization procedure is equivalent to the Stratonovich interpretation, $\alpha=1/2$~\cite{vanKampen,Zinn-Justin}. However, in other applications, 
like chemical Langevin equations~\cite{vanKampen} or econometric problems~\cite{Mantegna,Bouchaud},  
the noise can be considered principally white, since it could be a reduction of jump-like or Poisson like processes. 
In such cases, the It\^o interpretation ($\alpha=0$)
should be more suitable. Hence, the interpretation of equation~(\ref{eq.Langevin}) depends on the physics behind 
a particular application. 
Once the interpretation is fixed, the stochastic dynamics is unambiguously defined. Summarizing, in order to 
completely define the stochastic 
process described by equation~(\ref{eq.Langevin}) we need to fix a couple of
functions $(f,g)$ and the parameter $0\le\alpha\le 1$.

%%%%%%%%%%%%%%%%%%%%%%%%%%%%%%%%%%%%%%%%%%%%%%%%%%%%%%%%%%%%%
\section{Equilibrium  distribution and time reversal}
\label{TimeReversal}
%%%%%%%%%%%%%%%%%%%%%%%%%%%%%%%%%%%%%%%%%%%%%%%%%%%%%%%%%%%%%

The $\alpha$-prescription introduces some difficulties in the proper definition of 
time reversal evolution that we describe in this section. 

%%%%%%%%%%%%%%%%%%%%%%%%%%%%%%%%%%%%%%%%%%%%%%%
\subsection{Fokker-Planck approach}
Using equation~(\ref{eq.Langevin}) and the $\alpha$-prescription, it is
immediate to obtain~\cite{Lubensky2007} a Fokker-Planck equation for the
probability distribution $P(x,t)$:

\begin{equation}
\frac{\partial P(x,t)}{\partial t}=\frac{\partial}{\partial x}\left\{ -f-\alpha gg'\right\}P(x,t)+
\frac{1}{2}\frac{\partial^2}{\partial x^2}\left(g^2 P(x,t)\right),
\label{eq.FP}
\end{equation}
where $g'=dg/dx$.
This equation can be cast into a  continuity equation, 
\begin{equation}
\frac{\partial P(x,t)}{\partial t}+\frac{\partial J(x,t)}{\partial x}=0,
\label{eq.Continuity}
\end{equation} 
whith the probability current given by
\begin{equation}
J(x,t)= \left[f(x)-(1-\alpha) g(x)g'(x)\right] P(x,t)- \frac{1}{2}g^2(x) \frac{\partial P(x,t)}{\partial x}.
\label{eq.J(x,t)}
\end{equation} 
Then, the probability distribution $P(x,t)$ is given by the solution of the Fokker-Planck equation~(\ref{eq.FP})
supplemented with the initial condition $P(x,0)=P_{\rm in} (x)$ and two
boundary conditions. We can fix, for instance, the values of $J(x,t)$
at two boundary points $(x_i,x_f)$, where eventually they can be taken
as $\pm \infty$. However, due to the fact that
$P(x,t)$ is normalized to one, for any value of $t$, then $J(x_i,t)=J(x_f,t)$. 

We suppose that, at long times, the probability rapidly converges to a steady
state $P^S(x)$,  given by 
\begin{equation}
P^S(x)=\lim_{t\to \infty}P(x,t)=N\; e^{-U(x)},
\end{equation} 
with the normalization constant $N^{-1}=\int_{-\infty}^{\infty} dx\; e^{-U(x)}$. In this state, the stationary current gets the form 
\begin{equation}
J^S(x)= N e^{-U(x)}\;\left(f(x)-(1-\alpha) g(x)g'(x)+\frac{1}{2}g^2(x) \frac{dU(x)}{dx}\right)
\label{eq.JS}
\end{equation} 
and the stationary Fokker-Planck equation acquires the simpler form 
\begin{equation}
\frac{dJ^S(x)}{dx}=0,
\label{eq.stationaryFP}
\end{equation} 
or, simply, $J^S(x)=\bar J= \mbox{constant}$. 

Thus, for a given value of $\bar J$, we could have a stationary state
characterized by the function  $U_{\bar{J}}(x)$.
The stationary probability density $P^S_{\bar J}(x)=N_J\; e^{-U_{\bar J}(x)}$ does not represent,  in general, an equilibrium state. An  equilibrium state is characterized by 
a net zero stationary current ($\bar J=0$). When $\bar J\neq 0$, although the
current is already conserved, there is a stationary probability flux,
characterizing an 
out-of-equilibrium regime.  In most physical applications,  the diffusion
function $g(x)$ is a  polynomially growing
function for large $x$.
In these cases (and for a single variable), the only possible solution is the
equilibrium one  $P_{\rm eq}=P^S_{0}(x)=N_0\; e^{-U_{0}(x)}$. However, for
exponentially growing functions, out-of-equilibrium steady state solutions are
possible. 

Let us focus on the equilibrium state, defined as the solution of the
stationary Fokker-Planck equation with zero current probability. Thus,
for $\bar J=0$, there is an obvious solution of
equations~(\ref{eq.JS}) and (\ref{eq.stationaryFP}), given by 
\begin{equation}
U_{\rm eq}(x)=-2 \int^x \frac{f(x')}{g^2(x')} dx'+ (1-\alpha) \ln g^2(x).
\label{eq.equilibriumpotential}
\end{equation}
Therefore, given the functions $f(x)$ and $g(x)$, supplemented by a value of
$0\le\alpha\le1$, 
the probability density tends at long times to the equilibrium distribution 
\begin{equation}
P_{\rm eq}(x)= N e^{-U_{\rm eq}(x)},
\label{eq.Peq}
\end{equation} 
where $U_{\rm eq}(x)$ is given by equation~(\ref{eq.equilibriumpotential}), 
provided $e^{-U_{\rm eq}(x)}$ is integrable to define the normalization constant $N$.

In the case that equation~(\ref{eq.Langevin}) models a conservative physical
system, {\em i.\ e.\ }, if the otherwise deterministic system is
characterized by an energy potential $V(x)$,  
the drift function
$f(x)$ does not depend explicitly on time and is given by 
\begin{equation}
f(x)=-\frac{1}{2}g^2(x) \frac{dV(x)}{dx}.
\label{eq.conservative}
\end{equation}
This equation relates fluctuations (given by $g(x)$) with dissipation,
therefore it can be considered as a local generalization of Einstein relation
for Brownian motion. Using this expression, the
equilibrium
``potential''  has the form: 
\begin{equation}
U_{\rm eq}(x)= V(x)+(1-\alpha)\ln g^2(x).
\label{eq.U}
\end{equation} 
As expected, the equilibrium distribution depends, not only on the given functions $(f(x),g(x))$ or, alternatively, on  $(V(x),g(x))$, 
but also on the value of the $\alpha$-prescription which defines the Wiener integral. If equation~(\ref{eq.Langevin}) is used to model a physical system, which is expected to 
asymptotically converge to thermodynamic equilibrium ($U_{\rm eq}(x)= V(x)$), the only possible choice is  the H\"anggi-Klimontovich interpretation $\alpha=1$; 
for this reason, the election of $\alpha=1$ is sometimes called ``thermal
prescription''.
Recently, an experimental evaluation of $\alpha$ was
reported~\cite{Volpe2010,Volpe2011} in the study of a falling colloidal
particle near a wall. In this work, the value of $\alpha=1$ was recognized as
the proper one to make the correctly identification of the microscopic forces
acting on the colloidal particle.  
With any other choice,  the equilibrium potential $U_{\rm eq}(x)\neq V(x)$,
including the usual  It\^o ($\alpha=0$) and Stratonovich 
($\alpha=1/2$) prescriptions. 

Except for the Stratonovich prescription $\alpha=1/2$, each
value of $\alpha$ is associated with different calculus rules. For this reason,
in most physical literature, Stratonovich  prescription is preferred, because it enables ``normal'' rules of derivation and integration.
Despite the H\"anggi-Klimontovich convention being the only one which guarantees the convergence to the termodynamic equilibrium, calculations can be particularly cumbersome in this interpretation.
 However, if we prefer to
work in any other different fixed prescription, and we insist in modeling a physical system which converges
to Boltzmann's equilibrium, 
we need to modify equation~(\ref{eq.Langevin}). It is done by adding a ``spurious drift
force''~\cite{Lubensky2007} which cancels the last term of equation~ (\ref{eq.U}).
That is, if we define a new $\alpha$-dependent ``potential'' $\tilde
V_\alpha=V-(1-\alpha)\ln g^2(x)$, then, the equilibrium distribution  $U_{\rm
eq}(x)=V(x)$, for any value of  $\alpha$. 

In this work, we will not adopt this point of view. 
We will consider a stochastic process completely defined  by fixing  $(f(x),
g(x), \alpha)$ in eq.
(\ref{eq.Langevin}). Thus, the probability distribution
$P(x,t)$ flows, at long times,  
to an $\alpha$-dependent equilibrium distribution given 
by equation~(\ref{eq.equilibriumpotential}) or alternatively
equation~(\ref{eq.U}) .  In this way, we are able to study,
in a unique formalism, model systems with general equilibrium distributions that
go from Boltzmann thermal equilibrium 
to power-law distributions. A simple example of the latter case is to consider
a pure noisy system with $V(x)=0$. In that case, the equilibrium distribution
is $P_{\rm eq}\sim 1/g^{2(1-\alpha)}(x)$.

We will show that this set of distributions, characterized by asymptotically
zero current probability, is consistent with all the usual properties 
of equilibrium evolution, like, for instance, detailed balance and entropy
production.

%%%%%%%%%%%%%%%%%%%%%%%%%%%%%%%%%%%%%%%
\subsection{Time reversal}

Consider, for instance,  the Langevin equation~(\ref{eq.Langevin}), in the simplest case of $f(x)=0$.
Naively, we could compute the backward evolution of the stochastic variable
$x(t)$ by just changing the sign on the velocity  $\dot x\to -\dot x$, in such
a way that, for a given noise configuration,  the particle turns back on its
own feet. However, the situation is not so simple. Let us look more closely.
Consider a time interval $(t, t+\Delta t)$. The forward evolution is obtained
by integrating equation (\ref{eq.Langevin}) between the initial and final times
$t$ and  $t+\Delta t$ respectively,   
\begin{equation}
x(t+\Delta t )- x(t)=\int_t^{t+\Delta t} \frac{dx(t')}{dt'} dt'= \int_t^{t+\Delta t}g(x(t'))\zeta(t') dt',
\label{eq.fwevolution}
\end{equation}
while the backward evolution  is simply obtained by changing the initial
and final integration limits,  
\begin{equation}
\bar x(t)-\bar x(t+\Delta t )=\int^t_{t+\Delta t} \frac{d\bar x(t')}{dt'} dt'=
\int^t_{t+\Delta t}g(\bar x(t'))\zeta(t') dt',
\label{eq.bwevolution}
\end{equation}
where we use the notation $\bar x$, just to differentiate backward from forward
evolution.
If the integrals were ``normal'' integrals, we could use the trivial property $\int_a^b =-\int_b^a$. In that case, there would be no 
need to differentiate backward and forward variables, since
equations~(\ref{eq.fwevolution}) and~(\ref{eq.bwevolution}) would be the same 
equation and $x=\bar x$. 
However, the integrals are Wiener integrals and need to be carefully defined.
As we saw in the last section, 
\begin{equation}
\int_t^{t+\Delta t}    g(x(t))\;  dW(t)=\lim _{n\to\infty} \sum_{j=1}^n  g(x(\tau_j))(W(t_{j+1})-W(t_j)),
\label{eq.Wienerfw}
\end{equation}
with  
\begin{equation}
g(x(\tau_j))=g((1-\alpha)x(t_j)+\alpha x(t_{j+1})), \ \ \ 0\le \alpha \le 1. 
%g(x(\tau_j))=g((1-\alpha)x(t_j)+\alpha x(t_{j+1}))\mbox{~~ with~~} 0\le \alpha \le 1. 
\label{eq.prescriptionfw}
\end{equation}

The time reversal integral is obtained by changing $t_j\leftrightarrow t_{j+1}$
in equation~(\ref{eq.Wienerfw}). The important point is  that 
$g(x(\tau_j))$ also depends on $(t_j,t_{j+1})$.
Then,
\begin{equation}
\int^t_{t+\Delta t}  g(x(t))\;  dW(t)=\lim _{n\to\infty} \sum_{j=1}^n  \bar g(x(\tau_j))(W(t_{j})-W(t_{j+1})),
\label{eq.Wienerbw}
\end{equation}
with  
\begin{equation}
\bar g(x(\tau_j))=g(\alpha x(t_j)+(1-\alpha) x(t_{j+1})), \ \ \ 0\le \alpha \le 1. 
\label{eq.prescriptionbw}
\end{equation}
where $\bar g(x)$ was obtained from equation (\ref{eq.prescriptionfw}), by
replacing $t_j\leftrightarrow t_{j+1}$ or, equivalently, $\alpha\to (1-\alpha)$.
Therefore, the time reversed stochastic evolution is characterized by the
transformations 
\begin{equation}
x(t)\to x(-t)  ~~~~~\mbox{and}~~~~~~~\alpha \leftrightarrow (1-\alpha).
\end{equation}

\begin{figure}
\centering
\includegraphics[height=6.5cm]{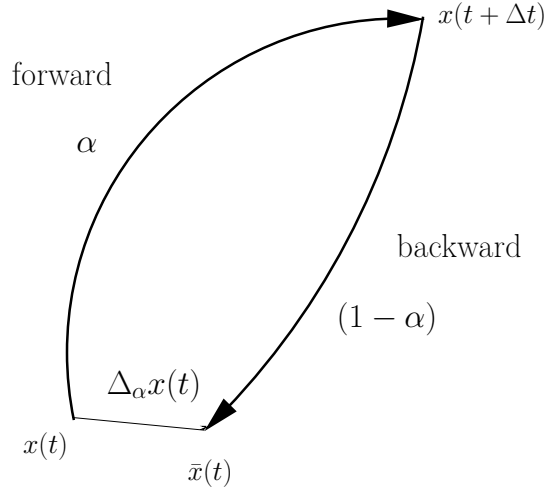}
\caption{Sketch of a ``time-loop''  evolution of Langevin equation
(\ref{eq.Langevin}). The forward evolution $t\to t+\Delta t$ is performed with a
prescription $\alpha$,  while the backward trajectory $t+\Delta t \to t$ evolves
with the time reversal conjugate prescription $(1-\alpha)$. Except for
the Stratonovich convention  $\alpha=1/2$, $\Delta x(t)_\alpha\neq 0$.}
 \label{timereversal}
 \end{figure}

In this sense, we say that the prescription $(1-\alpha)$ is the time reversal
conjugate (TRC) of $\alpha$. Then, the post-point H\"anggi-Klimontovich
interpretation is the  TRC of the pre-point It\^o one, and vice versa. The only
{\em time reversal invariant prescription} is the Stratonovich one,
$\alpha=1/2$. 
This means that, except for the Stratonovich case, the backward and forward
stochastic paths do not have the same end points. This is illustrated in figure
(\ref{timereversal}), where we compute a ``time-loop'' evolution. Consider we
want to compute the evolution of the system starting at $x(t)$, going forward a
time interval $\Delta t$ and, then, turning back in time the same interval
$-\Delta t$.

The forward  path in figure  (\ref{timereversal}) can be computed from equation
(\ref{eq.fwevolution}) considering $x(t)$ as an initial condition,
\begin{equation}
x(t+\Delta t )=x(t)+ \int_t^{t+\Delta t}g(x(t'))\zeta(t') dt'.
\label{eq.path1}
\end{equation}
Then, the backward path is computed from equation~(\ref{eq.bwevolution})
considering $x(t+\Delta t)$ as an initial condition
\begin{equation}
\bar x(t)=x(t+\Delta t )+ \int^t_{t+\Delta t}g(x(t'))\zeta(t') dt'.
\label{eq.path2}
\end{equation}
Replacing equation~(\ref{eq.path1}) into~(\ref{eq.path2}),
\begin{equation}
\Delta_\alpha x(t)\equiv \bar x(t)-x(t)=\int_t^{t+\Delta t}g(x(t'))\zeta(t') dt'+  \int^t_{t+\Delta t}g(x(t'))\zeta(t') dt',
\label{eq.deltax}
\end{equation}
and using equations~(\ref{eq.Wienerfw}) and~(\ref{eq.Wienerbw}) we find
\begin{equation}
\Delta_\alpha x(t)=\lim _{n\to\infty} \sum_{j=1}^n  	\left[g(x(\tau_j))-\bar g(x(\tau_j))\right]\left(W(t_{j+1})-W(t_j)\right).
\label{eq.loop}
\end{equation}
We see that, in general, $\Delta_\alpha x(t)\neq 0$ since the backward and
forward evolutions develop with different dual prescriptions.  In the
Stratonovich 
case, $\alpha=1/2$, $\bar g(x) = g(x)$ and $\Delta x_{1/2}(t)=0$.

%%%%%%%%%%%%%%%%%%%%%%%%
%%%%%%%%%%%%%%%%%%%%%%%%

This fact is reflected in the $\alpha$-dependence of the equilibrium stationary state $U_{eq}(x)$ (equation~(\ref{eq.equilibriumpotential}) or  equation~(\ref{eq.U})). While  the stationary state  in the forward evolution is $U_{fw}(x)=V(x)+(1-\alpha)\ln g^2(x)$, 
the stationary state reached by the backward  one  is given by 
$U_{bw}(x)=V(x)+\alpha\ln g^2(x)$, and (except for the Stratonovich
prescription) $U_{fw}\neq U_{bw}$. Thus, this is not the backward evolution we
are interested in. The key point is to determine the backward stochastic
process which, at long times, converges to the same equilibrium
distribution as the forward one. To obtain it we write the time reversed 
Fokker-Planck equation as
\begin{equation}
\frac{\partial \hat P(x,t)}{\partial t}=\frac{\partial}{\partial x}\left\{ \hat f+(1-\alpha) gg'\right\}\hat P(x,t)-
\frac{1}{2}\frac{\partial^2}{\partial x^2}\left(g^2 \hat P(x,t)\right),
\label{eq.FPBW}
\end{equation}
where we made the substitutions $x(t)\to x(-t)$, $\alpha\to (1-\alpha)$ and
$f\to \hat f$ in equation~(\ref{eq.FP}). In equation~(\ref{eq.FPBW}), $\hat
P(x,t)$ represents the backward probability distribution of the event $(x,t)$.
Now, we need to specify the drift $\hat f(x)$ which produces a backward
evolution that asymptotically converges to the equilibrium distribution
$P_{eq}=N
e^{-U_{eq}}$, with $U_{eq}(x)$ given by equation~(\ref{eq.U}). Setting $\partial
\hat P(x,t)/\partial t=0$ and  $P_{eq}=N e^{-U_{eq}}$, we solve
equation~(\ref{eq.FPBW}) for $\hat f$, obtaining
\begin{equation}
\hat f(x)= f(x) -\left(1-2\alpha\right)\; g(x) g'(x).
\end{equation} 

Therefore, the time reversal transformation ${\cal T}$ which makes physical sense, meaning that it 
produces a backward evolution which  converges to a unique equilibrium distribution, is  given by  

\begin{equation}
{\cal T} = \left\{
\begin{array}{lcl}
x(t) &\to & x(-t)   \\   & & \\
\alpha &\to & (1-\alpha) \\ & & \\
f &\to&  f -\left(1-2\alpha\right)\; g g'
\end{array}
\right.
\label{eq.T}
\end{equation}

Notice that, since $1-2\alpha$ is odd under the transformation $\alpha\to 1-\alpha$, the time reversal operator defined in this way satisfies 
${\cal T}^2=I$, as it should be. As we have previously stressed, in the
Stratonovich prescription $\hat f= f$ and $\alpha=1-\alpha$, so, the time
reversal operator ${\cal T}$ simply corresponds to change $x(t)\to x(-t)$. In 
any other prescription the definition of ${\cal T}$ is more involved, given by
equation~(\ref{eq.T}). Although we have defined the time reversal
transformation to match the correct equilibrium distribution in the long time
limit $\pm\infty$, this transformation also gives the correct answer in the
stationary out-of-equilibrium case $\bar J\neq 0$. In this case, the stationary
forward and backward probabilities are different, and they are related by $\bar
J \to -\bar J$. That is, $P^{bw}_{\bar J}(x)=P^{fw}_{-\bar J}(x)$ as can be
easily seen by applying the transformation (\ref{eq.T}) to the Fokker-Planck
equation (\ref{eq.Continuity}) and (\ref{eq.J(x,t)}).

%%%%%%%%%%%%%%%%%%%%%%%%%%%%%%%%%%%%%%%%%%%%%%%%
\section{Detailed balance, microscopic reversibility  and entropy production}
\label{DetailedBalance}

In this section we check that the concept of  $\alpha$-dependent equilibrium  and the time reversal transformation introduced in the previous sections
are consistent with the usual concepts of detailed balance, microscopic reversibility and entropy production. 
To do this, we use the path integral formalism, which turns out to be very convenient for handling stochastic trajectories.  

%%%%%%%%%%%%%%%%%%%%%
\subsection{Path Integral formulation}
\label{pathintegral}
The transition probability $P(x_f, t_f|x_i, t_i)$ represents the conditional
probability of finding the system in the state $x_f$ at time $t_f$ provided the
system was in the state $x_i$ at time $t_i$. In the path integral approach, it
can be written as \cite{arenas2010}

\begin{equation}
P(x_f, t_f|x_i, t_i)=
\int {\cal D}x\; {\det}^{-1}(g)\;  e^{-S[x]} \; ,
\label{eq.Pfw}
\end{equation}
with the boundary conditions $x(t_i)=x_i$ and $x(t_f)=x_f$ 
and the ``action'' $S[x]$ given by
\begin{equation}
 S = \int_{t_i}^{t_f} dt \left\lbrace\frac{1}{2g^2} \left[ \frac{dx}{dt} - f +
\alpha g g'\right]^2+\alpha f' \right\rbrace .
\label{eq.Sfw}
\end{equation}

The probability for the time reversal process $\hat P(x_i, t_f|x_f, t_i)$ is obtained by applying the time reversal transformation ${\cal T}$ of equation~(\ref{eq.T}) to the forward probability, equations~(\ref{eq.Pfw}) and~(\ref{eq.Sfw}),

\begin{equation}
\hat P(x_i, t_f|x_f, t_i)=
\int {\cal D}x\; {\det}^{-1}(g)\;  e^{-\hat S[x]} \; ,
\label{eq.Pbw}
\end{equation}
with the boundary conditions $x(t_i)=x_f$ and $x(t_f)=x_i$  and the time reversed ``action'' $\hat S[x]={\cal T} S[x]$.

In order to compute $\hat S$, it is interesting to write $S[x]$ in another form
to make explicit its properties under the time reversal transformation. 
Expanding the integrand of equation~(\ref{eq.Sfw}) and considering a
conservative system $f=-(g^2/2) V'$ (equation (\ref{eq.conservative})) we find 
\begin{eqnarray}
S = \int_{t_i}^{t_f} dt &&\left\lbrace
\frac{1}{2g^2}\left(\frac{dx}{dt}\right)^2 + \frac{1}{2g^2} \left( \frac{1}{2}g^2 V'+\alpha g g'\right)^2-\alpha \frac{1}{2}\left(g^2 V'\right)'
 +\right. \nonumber \\  & & \nonumber \\
&+& \left. \frac{1}{2}  \frac{d\left[V+\alpha\ln g^2\right]}{dx} \; \frac{dx}{dt}
\right\rbrace.
\label{eq.SfwV}
\end{eqnarray}
The last term can be transformed in a total time derivative. However, for
arbitrary values of $\alpha$, we need to apply the proper chain rule, since 
usual calculations rules are valid only for the Stratonovich convention $\alpha=1/2$. In the general case, for an arbitrary differentiable function of a random variable $Y(x(t))$,  the chain rule reads
\begin{equation}
\frac{dY(x(t))}{dt}=\frac{\partial Y}{\partial x}\; \frac{dx}{dt} +\frac{(1-2\alpha)}{2}\; \frac{\partial^2 Y}{\partial x^2}\; g^2.
\label{eq.chainrule}
\end{equation} 
Clearly, for $\alpha=1/2$, equation~(\ref{eq.chainrule}) is the usual chain rule. For $\alpha=0$, this formula is known as the \emph{It\^o formula}. 
For the H\"anggi-Klimontovich  prescription, $\alpha=1$, the latter differentiation rule differs from the It\^o formula just in the sign of the last term. This is because both  prescriptions are time reversal conjugate. In fact, the second derivative term is odd under $\alpha\to (1-\alpha)$, as it should be, since all terms in (\ref{eq.chainrule}) should be odd under time reversal. 

Applying equation~(\ref{eq.chainrule}) to the last term of equation~(\ref{eq.SfwV}), we find
\begin{equation}
S= \frac{1}{2}\left.\left(V+\alpha\;\ln g^2\right)\right|_{t_i}^{t_f}+ \tilde S,
\label{eq.SST}
\end{equation}
where 
\begin{eqnarray}
\tilde S = \int_{t_i}^{t_f} dt && \left\lbrace
\frac{1}{2g^2}\left(\frac{dx}{dt}\right)^2 + \frac{1}{8}g^2 (V')^2
-\frac{1}{2}\alpha g g' V' -\frac{1}{4} g^2 V''- \right. \nonumber \\  & & \nonumber \\ 
&-&\left. \alpha\left(\frac{1-2\alpha}{2}\right) g g'' + \frac{1}{2}\alpha(1-\alpha) (g')^2  
\right\rbrace.
\label{eq.Stilde}
\end{eqnarray}

The most important property which results from equations~(\ref{eq.SST})
and~(\ref{eq.Stilde}) is that $\tilde S$ is {\em time reversal invariant},  
$
\tilde S= {\cal T} \tilde S
$. 
Therefore, by using equation (\ref{eq.SST}), the time reversed action $\hat S$
can be easily determined:  
\begin{equation}
\hat S={\cal T} S= -\frac{1}{2}\left.\left(V+(2-3\alpha)\;\ln g^2\right)\right|_{t_i}^{t_f}+ \tilde S
\label{eq.Sbw}
\end{equation}
Thus, for an arbitrary  $\alpha$, the variation of the action $S[x]$
under a time reversal transformation is just a total derivative term. 
Using equations~(\ref{eq.SST}) and~(\ref{eq.Sbw}), it is immediate to show that 
\begin{equation}
S-\hat S= U_{\rm eq}(x_f)-U_{\rm eq}(x_i).
\label{eq.STS}
\end{equation}
with $U_{\rm eq}(x)$ given by equation~(\ref{eq.U}).

%%%%%%%%%%%%%%%%%%%%%%%%%%%%%%%%%%%%%%%%%%%%%%%%
\subsection{Detailed balance}
One of the  consequences of the time reversal operator~(\ref{eq.T})
is that the conditional probabilities $P\neq \hat P$. However, using
equations~(\ref{eq.Pfw}),~(\ref{eq.Pbw}),~(\ref{eq.SST}) and~(\ref{eq.Sbw}), it
is simple to show that 
\begin{equation}
\frac{P(x_f, t_f|x_i, t_i)}{\hat P(x_i, t_f|x_f, t_i)}=e^{-\left[V+(1-\alpha)\ln g^2\right]_{t_i}^{t_f}},
\end{equation}
satisfying the detailed balance relation 
\begin{equation}
P_{\rm eq}(x_i)\; P(x_f, t_f|x_i, t_i) = \hat P(x_f, t_i|x_i, t_f)\;P_{\rm eq}(x_f),
\label{eq.detailedbalance}
\end{equation}
with $P_{\rm eq}(x)$ given by equations~(\ref{eq.Peq}) and~(\ref{eq.U}).

So, we have shown that forward and backward conditional probabilities,
with the definition of equilibrium state given by the effective potential of
equation (\ref{eq.U}), satisfy the detailed balance relation,
equation~(\ref{eq.detailedbalance}), for arbitrary values of $0\le\alpha\le 1$.
In the special case of $\alpha=1/2$, $P=\hat P$, and
equation~(\ref{eq.detailedbalance}) is the usual detailed balance relation.

%%%%%%%%%%%%%%%%%%%%%%%%%%%%%%%%%%%%%%%%%%%%%%%%%
\subsection{Microscopic reversibility and entropy production}  

It is not difficult to analyze the previous results from the perspective of
stochas\-tic ther\-mo\-dy\-na\-mics~\cite{sasa2001}.
The detailed balance relation~(\ref{eq.detailedbalance}) represents a connection between a transition probability and its time reversed one.  Moreover, the transition probability was obtained by integrating over all 
possible stochastic trajectories.   However, there is also a relation between
the probabilities associated with each individual stochastic trajectory,  that
is usually called microscopic reversibility~\cite{crooks1999}. 
To show  up this relation  explicitly, we can assign to  each stochastic trajectory $x(t)$,  with end points $(x_i,t_i)$ and $(x_f,t_f)$, a weight 
\begin{equation}
{\cal P}(x(t)|x_it_i,x_ft_f)=  {\det}^{-1}(g)\;  e^{-S[x]} \; ,
\label{eq.Pxtfw}
\end{equation}
The heat dissipated into the environment $Q$, and thus, the increase of entropy in the medium $\Delta s_m$ associated 
with one specific trajectory, is given by~\cite{crooks2000} 
\begin{equation}
\Delta s_m=\beta Q[x(t)]=\ln\left[\frac{{\cal P}(x(t)|x_it_i,x_ft_f)}{\hat {\cal P}(x(t)|x_it_f,x_ft_i)}\right].
\label{eq.Q}	
\end{equation}
Hence, for each stochastic trajectory beginning in a state with initial distribution $p(x_i)$ and ending with a different distribution 
$p(x_f)$, the total entropy production is
\begin{equation}
\Delta s= \ln p(x_i)-\ln p(x_f)-\beta Q
\end{equation}
or, using equation~(\ref{eq.STS}),
\begin{equation}
\Delta s= \ln p(x_i)-\ln p(x_f)- U_{\rm eq}(x_f)+U_{\rm eq}(x_i).
\label{eq.Deltas}
\end{equation}
Thus, in the absence of an explicit time dependent driving force, the stochastic entropy is a state function which depends only on the initial and final states.  Moreover, if we prepare the initial state in equilibrium $p(x_i)=P_{\rm eq}(x_i)= N e^{-U_{\rm eq}(x_i)}$, we immediately conclude that $\Delta s=0$ for each stochastic trajectory and for arbitrary value of $\alpha$. 
The relevance of the potential $U_{\rm eq}$ in the entropy production  was
recently pointed out in~\cite{levkiselev2010} 
 for the Stratonovich prescription. Here, we generalize this concept to the
general case of the $\alpha$-prescription where the time reversed process is not
trivial. It is interesting to note that a similar expression to
(\ref{eq.Deltas}), however quite cumbersome, could be obtained for a
stationary out-of-equilibrium potential. The complication in this case is that
the stationary potential depends on $\bar J$, and the action does not transform
as  easily as in equation (\ref{eq.STS}) under time reversal, since $\bar J\to
-\bar J$. 

%%%%%%%%%%%%%%%%%%%%%%%%%%%%%%%%%%%%%%%%%%%%%%%%
\section{Correlation functions, Grassman variables and hidden symmetries}
\label{SUSY}
%%%%%%%%%%%%%%%%%%%%%%%%%%%%%%%%%%%%%

In order to push forward our analysis of equilibrium properties it is convenient to have a formalism to 
compute correlations and response functions.  However,  the action of equation (\ref{eq.SST}) and the 
difficulties introduced by the generalize It\^o calculus (eq. (\ref{eq.chainrule})) make the functional formalism in 
$x(t)$ quite cumbersome.  For this reason it is convenient to introduce auxiliary stochastic variables 
(a function $\varphi(t)$ and two Grassman functions $\xi(t), \bar\xi(t)$) extending in this way the functional 
space of stochastic functions.  The immediate consequence is a formalism that  does not depend on the explicit 
value of $\alpha$, simplifying its mathematical properties.  As a by-product,  hidden symmetries of the system 
can be put  in evidence. These symmetries impose non-perturbative constraints on  correlations functions and are 
related to equilibrium  properties of the system like, for instance, the fluctuation-dissipation theorem.   

%%%%%%%%%%%%%%%%%%%%%%%%%%%%%%%%%%%%%%%%%%%%%%%%
\subsection{Generating functional}
We are interested in the calculation of correlations of functions of the stochastic variable at equal or different times. 
For instance,
we could want to compute   $\langle A(x(t))\rangle$, where $A(x(t))$ is an arbitrary function of $x$, or two point 
functions $\langle A(x(t))B(x(t'))\rangle$.

In the path integral approach these mean values can be written as
\begin{equation}
\langle A(x(t))\rangle= \int dx_i dx_f P(x_i)\int_{x(t_i)=x_i} {\cal D}x\; {\det}^{-1}(g)\; A(x(t))\; e^{-S[x]} \; ,
\label{eq.A}
\end{equation}
where $S[x]$ is given by equation~(\ref{eq.Sfw}) and  $t_i<t<t_f$. This equation takes a simpler form if we consider that the system is 
prepared in equilibrium at $t=t_i$. In this case, $P(x_i)=P_{eq}(x_i)$. Assuming an ergodic evolution, we consider that 
\begin{equation}
P_{eq}(x_i)=\lim_{T\to\infty} P(x_i, t_i|x_{-T}, -T)
\label{eq.ergodic}
\end{equation} 
for an arbitrary value of $x_{-T}$.
Therefore, using equation (\ref{eq.A}) and equation (\ref{eq.ergodic}) we can write the mean value as 
\begin{equation}
\langle A(x(t))\rangle=\int{\cal D}x\; {\det}^{-1}(g)\; A(x(t))\; e^{-S[x]} \; ,
\end{equation}
where we have extended the time integration from $-\infty$ to $\infty$ and, now,  there are no more constraints in the 
functional integration. 
It is important to recall that this simpler expression is only possible because we sample the initial 
condition $x(t_i)$ with the equilibrium distribution  $P_{eq}(x_i)$. Provided this is satisfied, we can 
compute any mean value or correlation function from the generating functional 
\begin{equation}
Z(J)=\int{\cal D}x\; {\det}^{-1}(g) e^{-S[x]+\int_{-\infty}^{\infty}dt' J(t') x(t')} \; ,
\label{eq.ZS}
\end{equation}
where  $J(t)$ is a source with compact domain, that is, it adiabatically goes to zero away from  an  interval 
$(t_i, t_f)$ where we will compute the correlation functions. 

In this context, any total derivative term in $S[x]$ does not contribute to the dynamics of any observable. 
A total derivative term only  contributes to a constant prefactor of a correlation function. In this way, 
a system described by equation~(\ref{eq.ZS}) is automatically invariant under time reversal since $S$ and 
$\hat S={\cal T} S$ differ just in a total time derivative term. On the other hand, if we impose the constraint 
$x(+\infty)=x(-\infty)$, the action is truly time reversal invariant $S[x]=\hat S[x]=\tilde S$. Again, it is 
important to observe that this happens because we have prepared the system at equilibrium at some initial time 
$t_i$ and we are computing correlation functions at later times $t>t_i$. 
For any other initial distribution we need to go back to equation~(\ref{eq.A}) to compute mean values. 
 
%%%%%%%%%%%%%%%%%%%%%%%%%%%%%%%%%%%%%%%%%%%%%%%%%%%%%
\subsection{Time reversal and entropy production in  terms of Grassman variables}

Equation~(\ref{eq.ZS}) can be used to study general correlation and response functions of a white-noise 
multiplicative stochastic process whose dynamics is driven by equation~(\ref{eq.Langevin}). 
The $\alpha$-dependence of the final equilibrium distribution, the structure of $\tilde S$ and the fact that 
for general $\alpha$ the usual rules of calculus do not apply, make this method cumbersome. 
Nevertheless, we have shown~\cite{arenas2010} that, introducing an auxiliary
``bosonic'' variable $\varphi(t)$, and a couple of conjugate Grassman variables
$\bar\xi(t), \xi(t)$, we can avoid this inconvenience by alternatively using the
generating functional formalism in this extended space.
The generating functional can be cast in the following
form, 
\begin{equation}
 \mathit{Z}[J] = \int {\cal D}x {\cal D}\varphi {\cal D}\xi {\cal D}\bar{\xi} \ e^{-S[x,\varphi,\xi,\bar\xi] + \int dt J(t)x(t)} \ ,
\label{eq.funcionalgenerador}
\end{equation}
where the ``action'' $S$ is given by
\begin{eqnarray}
 S = \int_{-\infty}^{+\infty} dt && \left\lbrace  -\bar{\xi}(t)\frac{d}{dt}\xi(t) + f'(x)\bar{\xi}(t)\xi(t) +\frac{1}{2}\varphi(t)^2 g(x)^2 +
 \right.\nonumber \\ 
 & & + \left. i\varphi(t) \left[ \frac{dx}{dt} - f(x) + g(x)g'(x)\bar{\xi}(t)\xi(t)\right]\right\rbrace .
\label{eq.SGrassman}
\end{eqnarray}
This action in the extended functional space $(x(t), \varphi(t),\xi(t),\bar\xi(t))$ is completely equi\-va\-lent to the 
usual formalism presented in \S \ref{pathintegral}.  In fact,  functional integrating over Grassman variables and 
over the variable $\varphi$ we get the generating functional~(\ref{eq.ZS}), with the action given by 
equation~(\ref{eq.Sfw}). The advantage of this extended formulation is that it
does not depend explicitly of 
the $\alpha$-parameter and then, the calculus rules are the usual ones for any $\alpha$.   
Hence, all the complication of the stochastic calculus associated with the definition of the Wiener integral 
is now codified in the anti-commuting  Grassman  algebra, implying, for instance, that $\xi(t)^2=\bar\xi(t)^2=0$.
Of course, the parameter $\alpha$ reappears when we need to properly define
Grassman retarded Green's functions. In fact, equal-time correlations functions
are  ill-defined, forcing us to set $\langle\bar\xi(t)\xi(t)\rangle=\alpha$. 
One of the advantages of this procedure is that, while the Grassman variables
are not integrated, we can work out any calculation without explicitly indicate
an specific convention. This is particularly useful to study
Markovian white-noise processes,  where the equilibrium distribution depends on
$\alpha$. This
situation is very different from non-Markov processes~\cite{AronLeticia2010}, in
which $\langle\bar\xi(t)\xi(t)\rangle=0$.

The time reversal transformation, equation~(\ref{eq.T}), has a simpler form in terms of Grassman variables,
\begin{equation}
{\cal T} = \left\{
\begin{array}{lcl}
x(t) &\to  &  x(-t)  \\
\varphi(t) & \to  & \varphi(-t) -\frac{2i}{g^2}\; \dot x(-t) \\
\xi(t) &\to & \bar\eta(-t)   \\
\bar\xi(t) &\to &- \eta(-t)
\end{array}
\right.
\label{eq.TG}
\end{equation}
where $\eta, \bar\eta$ are the time reversed Grassman variables.  
There is a subtlety in these transformations which involves the extremal values
of the Grassman variables. 
While $\lim_{t\to \pm\infty} \bar\xi(t)\xi(t)=\alpha$, the time
reversed variables satisfy 
$\lim_{t\to \pm\infty} \bar\eta(t)\eta(t)=1-\alpha$. This fact  
has no importance at all when computing correlation functions. However, it is
relevant to analyze entropy production and other relations that depend on total
derivatives.

The real advantage of the Grassman representation is that  almost any important symmetry becomes a linear transformation. 
It is simple to check that, in this representation,  ${\cal T}^2 =I$ and 
\begin{equation}
S-\hat S= U_{eq}^f-U_{eq}^i+ \int_{-\infty}^{\infty} \ln g^2\; \frac{d~}{dt}\left(\bar\xi\xi\right) \; dt.
\end{equation}
Then, the entropy production $\Delta s$, associated with each trajectory, is given by 
\begin{equation}
\label{eq.entropy}
\Delta s=\int_{-\infty}^{\infty} \ln g^2\; \frac{d~}{dt}\left(\bar\xi\xi\right) \; dt.
\end{equation}
We see that, in the extended space of trajectories $(x,\varphi,\xi, \bar\xi)$, the entropy production associated with 
each trajectory is not zero. This is due to the apparently extra degrees of freedom introduced by the Grassman 
variables. On the other hand, upon functional integration we have that 
\begin{equation}
\frac{d~}{dt}\langle\bar\xi\xi\rangle=0,
\end{equation}
due to fermionic number conservation and the total entropy production is zero, as it should be in an 
equilibrium evolution.  This opens the interesting possibility of interpreting an entropy production
 as a spontaneous breaking of fermionic  phase (Gauge) invariance like, for instance, in superconducting 
phase transition. This effect cannot happen in our example of one stochastic variable, but in multiple 
variables systems~\cite{Puglisi2012}, there are quartic interactions
between Grassman variables opening this attractive possibility. 

%%%%%%%%%%%%%%%%%%%%%%%%
 \subsection{Linear response}
It is interesting to compute the linear response of the system to an external perturbation. 
To see this,  we slightly perturb the system out of equilibrium  
\begin{equation}
V(x)\to V(x) - h(t) x(t), 
 \label{eq.perturb}
\end{equation}
and compute the 
dynamic susceptibility 
\begin{equation}
 \chi(t,t') \equiv\left.\frac{\delta \langle x(t)\rangle_h}{\delta h(t')}\right|_{h=0}.
 \label{eq.chi}
 \end{equation}
Introducing~(\ref{eq.perturb}) into~(\ref{eq.SGrassman}) and computing equation~(\ref{eq.chi}) we find 
\begin{equation}
 \chi(t,t') =i \langle x(t), g(x')^2\varphi(t') \rangle - \langle x(t), g(x')g'(x')\xi(t')\bar\xi(t') \rangle
 \label{eq.response}
 \end{equation}
In the case of additive noise, where $g=\mbox{constant}$, the response function has the simpler form, 
$\chi(t,t') \sim \langle x(t), \varphi(t') \rangle $. For this reason, the auxiliary function $\varphi(t)$
is usually called the response variable.  However, for multiplicative  noise, the response is more complex since 
involves, not only correlations functions of $g(x)$, but also contributions from
the Grassman sector of the model. This is a direct consequence of multiplicative white-noise  since
 a variation of the ``external'' potential is
modifying the fluctuations properties of the system. 
This means that the susceptibility  explicitly depends on $\alpha$. 
From equation~(\ref{eq.response}), it is immediate to recognize a {\em ``natural
response variable''}
\begin{equation}
	\tilde\varphi(t)= g^2(x)\left(\varphi(t)+i \frac{d\ln g^2}{dx}\bar\xi(t)\xi(t)\right),
\label{eq.NaturalResponseVariable}
\end{equation}
in such a way that, 
\begin{equation}
 \chi(t,t') =i \langle x(t), i \tilde\varphi(t') \rangle
 \label{eq.NaturalResponse}
 \end{equation}
We will show that $\tilde\varphi(t)$ has a key role in the supersymmetric formulation of the model.

%%%%%%%%%%%%%%%%%%%%%%%%%%%%%%%%%%%%%%%%%%%%%%%%%%
\subsection{Supersymmetry and covariant (superfield) representation}

The path integral formalism is useful to make evident  symmetries of the stochastic process. For instance, 
the action given by eq.(\ref{eq.SGrassman}) is invariant under the transformation, 
\begin{eqnarray}
\delta x &=& \bar\lambda \xi\mbox{~~~~~,~~~~~~} \delta \xi=0 \ ,  \label{BRS1}\\
\delta \bar\xi &=& i\bar\lambda \varphi\mbox{~~~~,~~~~~~} \delta \varphi=0 \ , \label{BRS2} 
\end{eqnarray}
where $\bar\lambda$ is an anticommuting parameter.  This nilpotent transformation  ($\delta^2=0$) is the famous BRS~\cite{bechi1976} 
symmetry, discovered in the context of quantization of Gauge theories. In the present context, it simply enforces  probability 
conservation,  $Z(0)=1$. 

There is another set of important symmetries related with equilibrium properties which is called supersymmetry. 
To display it explicitly, it is convenient to work with the  natural response variable introduced in equation (\ref{eq.NaturalResponseVariable}).
Therefore, we make in equations (\ref{eq.funcionalgenerador}) and (\ref{eq.SGrassman}) the funcional change of variables $(x,\varphi,\xi,\bar\xi)\to (x,\tilde\varphi,\eta,\bar\eta)$, 
defined by  
\begin{eqnarray}
	\tilde\varphi(t)&=& g^2(x(t))\left(\varphi(t)+i \frac{d\ln g^2(x(t))}{dx}\bar\xi(t)\xi(t)\right) \\
             \eta(t) &=& g(x(t)) \xi(t)  \\
        \bar\eta(t)&=&g(x(t)) \bar\xi(t).
\end{eqnarray}
The last two transformations make  the Jacobian trivial,  
${\cal D}\varphi {\cal D}\xi {\cal D}\bar{\xi}={\cal D}\tilde\varphi {\cal D}\eta {\cal D}\bar{\eta}$, since ${\cal D}\tilde\varphi=\det^2(g){\cal D}\varphi$ and 
the fermionic measure reads ${\cal D}\eta {\cal D}\bar{\eta}=\det^{-2}(g){\cal D}\xi {\cal D}\bar{\xi}$.

The  generating functional is then given by 
\begin{equation}
 \mathit{Z}[J] = \int {\cal D}x {\cal D}\tilde\varphi {\cal D}\eta {\cal D}\bar{\eta} \ e^{-S[x,\tilde\varphi,\eta,\bar\eta] + \int dt J(t)x(t)} \ ,
\label{eq.Zeta}
\end{equation}
where the action $S[x,\tilde\varphi,\eta,\bar\eta]$ reads
\begin{eqnarray}
 S = \int_{-\infty}^{+\infty} dt &&\left\lbrace  -\bar{\eta}g^{-1}\frac{d}{dt}\left(g^{-1}\eta\right) + f'g^{-2}\bar{\eta}\eta +\frac{1}{2g^2}\tilde\varphi^2  +
 \right.\nonumber \\ 
& & \nonumber \\
 &+& \left. i g^{-2}\tilde\varphi \left[ \frac{dx}{dt} - f -\frac{g}{g'}\bar{\eta}\eta\right] + \frac{g'}{g^3}\left(\dot x-f\right)\bar\eta\eta\right\rbrace , 
\label{eq.Seta}
\end{eqnarray}
where we have drop-off function's arguments just to simplify notation. 

At first sight, this action written in the space $(x,\tilde\varphi,\eta,\bar\eta)$ seems to be more complex 
than the preceding one. However, these new physical variables can be 
rearranged in order to explicitly display SUSY. 

First, it is convenient to write $f(x)$ and $g(x)$ in terms of two ``potentials'', $V(x)$ and $\Gamma(x)$, defined 
as follows
\begin{eqnarray}
f(x)&=&-\frac{1}{2}g^2(x) \frac{dV(x)}{dx}, \\
g^{-1}(x)&=& \frac{d\Gamma(x)}{dx}. \label{eq.potentials}
\end{eqnarray}
Then,  we collect the variables $x,\tilde\varphi, \bar\eta,\eta$, in the definition of the superfield
\begin{equation}
\Phi(t,\theta,\bar\theta) = x(t)+ \bar \theta\eta(t)+ \bar\eta(t)\theta  +i \tilde\varphi(t) \bar\theta \theta.
\end{equation}
where we have introduced  two ``temporal''  Grassman variables $\theta$ and $\bar\theta$.

The generating functional, equation~(\ref{eq.Seta}), can be re-written in terms of $\Phi(t,\theta,\bar\theta)$ as
\begin{equation}
 \mathit{Z}[J] = \int {\cal D}\Phi \ e^{-\tilde S[\Phi] + \int dt d\theta d\bar \theta\; J(t,\theta,\bar\theta)\Phi(t,\theta,\bar\theta)} \ ,
\label{eq.GFPhi}
\end{equation}
where $\tilde S[\Phi]$ depends on the potentials $V(\Phi)$ and $\Gamma(\Phi)$ by 
\begin{equation}
\tilde S[\Phi] = \int dtd\theta d\bar\theta\; L(\Phi)=\int dtd\theta d\bar\theta\; \left\{\bar D\Gamma[\Phi] D\Gamma[\Phi]+\frac{1}{2} V[\Phi]\right\}.
\label{eq.SSUSY}
\end{equation}
We have defined the covariant derivatives 
\begin{equation}
\bar D=\frac{\partial~}{\partial\theta} \mbox{~~~,~~~} D = \frac{1}{2} \frac{\partial~}{\partial\bar\theta}-\theta\frac{\partial~}{\partial t} \ ,
\label{eq.cov_derivs}
\end{equation}
which satisfy  $D^2=\bar D^2=0$ and $\left\{D,\bar D\right\}=-\frac{\partial~}{\partial t}$.
The action~(\ref{eq.SSUSY}), when written in components form, is completely equivalent to equation~(\ref{eq.SGrassman}) or, 
after integration of the Grassman variables, to equation~(\ref{eq.Stilde}), as it is shown in the Appendix.

It is now immediate to verify that (\ref{eq.SSUSY}) is invariant under transformations of  the supersymmetry group 
\cite{Zinn-Justin}, whose generators are
\begin{equation}
 Q = \frac{\partial~}{\partial\bar\theta} \mbox{~~,~~} \bar Q = \frac{1}{2}\frac{\partial~}{\partial\theta}+\bar\theta\frac{\partial~}{\partial t}
\mbox{~~,~~} \left\{Q,\bar Q \right\}=\frac{\partial~}{\partial t} \ .
\label{eq.generators}
\end{equation}
The symplicity of the superfield representation resides in the fact that each SUSY transformation is a translation of 
the temporal variables 
$(t,\theta,\bar\theta)$.

Then, the invariance of the last term of equation~(\ref{eq.SSUSY}) is trivial, since $V(\Phi)$ does not depend 
explicitly on $(t,\theta,\bar\theta)$.
To demonstrate the invariance of the kinetic part of the action, it is useful to have in mind the graded algebra, 
\begin{eqnarray}
\lefteqn{
\left\{Q, D \right\}=\left\{Q, \bar D \right\}=\left\{\bar Q, D \right\}=\left\{\bar Q, \bar D \right\}=0 ,\label{eq.algebra}}  \\
Q^2 &=& \bar Q^2=0 \ .
\end{eqnarray}
Thus,
\begin{equation}
 \delta\left\{\bar D\Gamma[\Phi] D\Gamma[\Phi]\right\}=\bar D\delta\Gamma[\Phi] D\Gamma[\Phi]+\bar D\Gamma[\Phi] D\delta\Gamma[\Phi].
\end{equation}
On the other hand, $\delta\Gamma[\Phi]=\epsilon Q_\beta \Gamma[\Phi]$, where $\epsilon$ is an arbitrary Grassman parameter and $Q_\beta$ may represent 
$Q$ or $\bar Q$ defined in equation~(\ref{eq.generators}).  In this way, 
\begin{equation}
\delta\left\{\bar D\Gamma[\Phi] D\Gamma[\Phi]\right\}=\epsilon\left\{-\bar D\left( Q_\beta\Gamma[\Phi]\right) D\Gamma[\Phi]+\bar D\Gamma[\Phi]
 D\left(Q_\beta\Gamma[\Phi]\right)
\right\}
\end{equation}
Finally, using the algebra~(\ref{eq.algebra}), we find, 
\begin{equation}
\delta\left\{\bar D\Gamma[\Phi] D\Gamma[\Phi]\right\}=\epsilon Q_\beta\left\{\bar D\Gamma[\Phi] D\Gamma[\Phi]\right\},
\end{equation}
which is a total derivative term, demonstrating in this way, the invariance of
$\tilde S$.

For additive processes, the diffusive potential is linear, $\Gamma(x)\sim x$, and eq.(\ref{eq.SSUSY}) 
reduces to the usual action defined with a single superpotential $V(\Phi)$. In this case, the tadpole theorem~\cite{arenas2010} 
guarantees that the stochastic evolution does not depend on $\alpha$. However, multiplicative processes (non-linear $\Gamma$) 
induce derivative couplings in the superfield. These couplings are responsible for the  $\alpha$-dependent evolution which leads 
to the equilibrium distribution of equation~(\ref{eq.U}).

%%%%%%%%%%%%%%%%%%%%%%%%%%%%%%%%%%%%%%%%%%%%%%%%%%%%%
\subsection{SUSY Ward-Takahashi identities and equilibrium}

The study of the symmetries of the action enables us to deduce  properties of equilibrium dynamics of the system. 
Each of the three generators of SUSY, $\{ Q, \bar Q, \partial_t\}$, imposes several non-perturbative constraints on correlation
functions. To obtain these constraints, we will use the Ward-Takahashi identities 
\begin{equation}
  \langle\Phi( t_1,\theta_1,\bar\theta_1)\Phi( t_2,\theta_2,\bar\theta_2)\rangle = \langle e^{\ \epsilon  G}\ \Phi( t_1,\theta_1,\bar\theta_1) e^{\ \epsilon  G}\ \Phi( t_2,\theta_2,\bar\theta_2) \rangle,
 \label{eq.WT_Ident}
\end{equation}
where $\epsilon$ is an arbitrary parameter and $G$ is one of the generators of SUSY. Note that, if $G$ is $Q$ or $\bar Q$, 
then $\epsilon$ is a 
Grassman parameter, 
while, if $G=\partial_t$, $\epsilon$ is a commutative parameter. 

The Ward-Takahashi identity  related with the generator $\partial_t$ reads, 
\begin{equation}
\langle\Phi( t_1,\theta_1,\bar\theta_1)\Phi( t_2,\theta_2,\bar\theta_2)\rangle =  \langle\Phi(t_1+\epsilon,
\theta_1,\bar\theta_1)\Phi(t_2+\epsilon,\theta_2,\bar\theta_2)\rangle,
\label{eq.partial_t}
\end{equation}
implying time-translation invariance as
\begin{equation}
(\partial_{t_1} + \partial_{t_2})\langle\Phi( t_1,\theta_1,\bar\theta_1)\Phi( t_2,\theta_2,\bar\theta_2)\rangle = 0.
 \label{eq.time_invariance}
\end{equation}
Then, any two-point correlation function depends on time differences $t_2-t_1$. This is a necessary condition for an equilibrium evolution.

On the other hand, the generator of $\bar\theta$ translations, $Q$, induces the  identity
\begin{equation}
\langle\Phi( t_1,\theta_1,\bar\theta_1)\Phi( t_2,\theta_2,\bar\theta_2)\rangle =  \langle\Phi(t_1,
\theta_1,\bar\theta_1 + \epsilon)\Phi(t_2+\epsilon,\theta_2,\bar\theta_2 + \epsilon)\rangle,
\label{eq.partial_theta}
\end{equation}
that, when written in components, imposes the relations
\begin{eqnarray}
\langle x(t_1) i\tilde \varphi(t_2)\rangle & = & \langle\bar\eta(t_2)\eta(t_1)\rangle,   \label{eq.linearResponse_ETA}\\
\langle \tilde \varphi(t_i) \tilde \varphi(t_2)\rangle & = & 0.
\end{eqnarray}
All components, involving only one of the Grassman variables, $\eta$ and $\bar \eta$, vanish because of fermionic 
number conservation.
The first equation relates the physical linear response of the system with the Grassman two-point correlation function. 
Since the physical response is causal, 
{\em i.\ e.\ },   $\langle x(t_1) i\tilde \varphi(t_2)\rangle=0$ for $t_1<t_2$, (see equation (\ref{eq.NaturalResponse})), 
we are forced to choose 
the retarded prescription in computing  Green's functions of Grassman variables. On the other hand, the equal-time Green 
function is not well defined, forcing 
us to adopt a prescription  $\langle\bar\eta(t)\eta(t)\rangle=\alpha g^2(x_\infty)$, where $x_\infty$ is an arbitrary 
initial value.  This presciption, when written 
in the original variables, is $\langle\bar\xi(t)\xi(t)\rangle=\alpha$, allowing us to identify $\alpha$ with the needed
prescription to define the Wiener integral. 
Notice that, in the case of non-Markovian processes, $\langle\bar\xi(t)\xi(t)\rangle=0$, without any ambiguity.

Finally, the invariance of the action generated by $\bar Q$, which arise from the identity
\begin{eqnarray}
\lefteqn{\langle\Phi( t_1,\theta_1,\bar\theta_1)\Phi( t_2,\theta_2,\bar\theta_2)\rangle  = } \nonumber \\ 
\langle\Phi(t_1+\epsilon \bar\theta_1,\theta_1+\epsilon/2,\bar\theta_1)\Phi(t_2+\epsilon \bar\theta_2,\theta_2+\epsilon/2,\bar\theta_2)\rangle,
\label{eq.PhiPhi}
\end{eqnarray} 
results in the non-perturbative constraints, 
\begin{eqnarray}
 \langle \partial_{t_1} x(t_1) x(t_2)\rangle  & = &  \frac{1}{2} \langle i\tilde \varphi(t_1) x(t_2)\rangle + \frac{1}{2} \langle \eta(t_1) \bar\eta(t_2)\rangle, \label{eq.WT-FDT} \\
 \langle \partial_{t_1} x(t_1) i\tilde \varphi(t_2)\rangle & = & \langle\eta(t_1)\partial_{t_2} \bar\eta(t_2)\rangle.
\end{eqnarray}
This constraints  imply another important equilibrium property, the fluctuation-di\-ssi\-pa\-tion theorem, which relates 
the spontaneous 
fluctuations of the system with its response to an external
perturbation. To explicitly see this,  we 
substitute relation~(\ref{eq.linearResponse_ETA}) into~(\ref{eq.WT-FDT}), obtaining
\begin{equation}
 \partial_{t} \langle x(t) x(t')\rangle  =  \frac{1}{2} \langle i\tilde \varphi(t) x(t')\rangle - \frac{1}{2} \langle x(t) i\tilde \varphi(t') \rangle. \label{eq.FDT1} 
\end{equation}
Using the equilibrium time translations invariance of the correlation function $ C(t,t') = \langle x(t) x(t')\rangle $, this identity becomes
\begin{equation}
  \frac{1}{2} (\partial_t - \partial_{t'}) C (t-t') = \frac{1}{2} {\chi(t'-t) - \chi(t-t')},
\end{equation}
which, bearing causality in mind, can be rewritten as
\begin{equation}
  \chi(t-t') = - (\partial_t - \partial_{t'}) C (t-t') \Theta(t-t'),
  \label{FDT}
\end{equation}
where $\Theta(t)$ is the Heaviside step function.
Equation~(\ref{FDT}) is a traditional form of the fluctuation-dissipation theorem. 
A more commonly used form of equation~(\ref{FDT}) is given in terms of the Fourier transform of the response function, 
\begin{equation}
	\langle x(\omega) x(-\omega)\rangle = \frac{\mathcal{I}m\ \chi(\omega)}{\omega}. 
 \label{eq.FT_FDT}
\end{equation}
Then, the fluctuation-disipation theorem can be understood as a direct consequence of supersymmetry. We have 
shown that, even thought the correlation 
functions and the responses depend on $\alpha$ in a multiplicative process, the fluctuation-dissipation 
theorem is satisfied for any value of $\alpha$, 
provided we correctly identify the equilibrium distribution, equation~(\ref{eq.U}) and
the time reversal transformation, equation~(\ref{eq.T}).
These physical concepts are mathematically codified in the invariance under
supersymmetry trans\-for\-ma\-tions, that we have explicitly shown in this
section.

It is very interesting to note that extensions of the
fluctuation-dissipation theorem, for out-of-equilibrium steady states, can be
formulated~\cite{Marconi2008}. A natural arising question is whether a
related SUSY exists for these out-of-equilibrium states. 

%%%%%%%%%%%%%%%%%%%%%%%%%%%%%%%%%%%%%%%%%%%%%%%%%%%%%%%%%%%%%%%%%%
\section{Discussion and conclusions}
\label{discussion}
%%%%%%%%%%%%%%%%%%%%%%%%%%%%%%%%%%%%%%%%%%%%%%%%%%%%%%%%%%%%%%%%%

We have presented a detailed study of equilibrium properties of a
{\em Markovian mul\-ti\-pli\-ca\-tive white-noise stochastic process}.  The stochastic
process was modeled in its simplest version of a single stochastic
variable, by means of the Langevin equation~(\ref{eq.Langevin}),  defined by a
drift term $f(x)$ and a diffusion function $g(x)$. We have completed the
definition of the stochastic differential equation by using the {\em
Generalized Stratonovich Convention}, characterized by a parameter
$0\le\alpha\le 1$.

The asymptotic equilibrium distribution was found by solving the stationary
Fokker-Planck equation, imposing an asymptotic zero current probability. This
na\-tu\-ra\-lly conduce to an equilibrium potential $U_{\rm eq}(x)$ (Eq.~(\ref{eq.U}))
which explicitly depends on $\alpha$. For $\alpha=1$ (H\"anggi-Klimontovich
interpretation), the  potential coincides with the classical
deterministic one, $V(x)$, in agreement with the Boltzmann distribution for
thermodynamical equilibrium. However, for any other convention, the equilibrium
potential is more general, including, for instance, power-law distributions. 
Thus, the definition of equilibrium in the stochastic dynamical sense not
necessarily coincides with the thermodynamical equilibrium concept. We
have shown that, even for this generalized definition of equilibrium, the
system satisfies the usual equilibrium properties such as detailed balance,
microscopic reversibility and the fluctuation-dissipation theorem.

Time reversal properties of the stochastic process are quite interesting. We
have shown that, if the forward stochastic trajectory evolves with a definite
value of $\alpha$, the time reversed trajectory evolves with the conjugated
prescription $(1-\alpha)$. Therefore, in order to have a unique equilibrium
distribution, the definition of time reversed stochastic process is given by
the transformation~(\ref{eq.T}), where we need to change not only the velocity
sign, but also the prescription and the drift force.  

 We have shown that, using the equilibrium potential $U_{\rm eq}$ and the time
reversal transformation of equation~(\ref{eq.T}), the stochastic process satisfies
detailed balance relations for any value of the parameter $\alpha$. It is
convenient to note that, in equation~(\ref{eq.detailedbalance}), the forward
transition probability is not equal to the backward one, except for the
Stratonovich convention $\alpha=1/2$, where  forward and backward 
trajectories coincide. We have also shown that, if the
 initial and final states of a finite time evolution are prepared in equilibrium,
the entropy production of each stochastic trajectory vanishes for arbitrary
prescriptions, verifying the related principle of microscopic
reversibility. 
In order to show these properties, we have used the path integral
representation of the stochastic process,  where the central object is the hole
stochastic trajectory. Also,  we were forced to intensively use an
$\alpha$-generalized It\^o calculus, summarized in the $\alpha$-dependent
``chain rule'' of equation~(\ref{eq.chainrule}).

Related with equilibrium properties, there are hidden symmetries in the
stochastic process that can be better analyzed in the path integral approach,
formulated in an extended functional space composed of commutative as well as
anti-commutative Grassman variables. The advantage of this formulation is that
it is prescription independent.  That is, ambiguities in the definition of the
multiplicative stochastic processes can be understood, not as a continuum limit
ambiguity, but as a necessity to define equal time Grassman Green's functions.
Then, provided the Grassman variables are not integrated out, it is not
necessary to fix any prescription. In this way, we can use ordinary calculus rules, 
but instead we need to work with
anti-commuting variables. In some sense, all the complexity of the It\^o
calculus is codified in the Grassman algebra in this formulation.

In this context,  we have computed the
entropy for each stochastic trajectory (in the extended space) and we have shown
that it is not zero, but proportional to a time derivative of the Grassman
variables density  $\bar\xi\xi$ (equation~(\ref{eq.entropy})). 
This is due to the extra degrees of freedom we have added 
to extend the functional space. However, integrating over the Grassman
variables, we immediately recognize that the entropy production is zero, as it
should be for an equilibrium evolution. Interestingly, this cancellation is due
to ``{\em fermionic number conservation}'', in other words, due to the invariance
under phase transformations. This allows us to speculate
about the very attractive possibility of interpreting an out-of-equilibrium
evolution, {\em i.\ e.\ }, a non-zero entropy production, as a consequence of a gauge
symmetry breaking, very similar with a superconducting transition. Of course,
this cannot happen in our present example, since we only have one stochastic
variable and, consequently, the Grassman variables are not-interacting. Nevertheless,
in multiple variable systems, quartic terms in the Grassman variables are
certainly possible.

Another interesting symmetry which appears in this formulation is supersymmetry (SUSY),
expressed as a linear transformation, mixing commuting as well as
anti-commuting variables. We have presented a supersymmetric covariant
representation of the stochastic process. While this representation was known
for additive processes  and multiplicative non-Markov  processes, the present
case of {\em Markovian multiplicative white noise} was quite difficult to treat due
to the different prescriptions available to define it, connected by a time reversal
transformation. In this representation, the non-trivial time reversal
transformation is represented by a simple linear transformation. The difficulty
of the time conjugated prescriptions are confined to boundary conditions,
irrelevant in the computation of correlation functions. 

Using this SUSY covariant formalism we have analyzed  two-point
Ward-Takahashi identities. One of them relates linear responses and
fluctuations. We have shown that, while the linear response function and
fluctuations are $\alpha$-dependent, they satisfy the fluctuation-dissipation
theorem for arbitrary values of the prescription
$\alpha$. Therefore, similar to the additive case, supersymmetry is a
consequence of equilibrium evolution, even thought we have different
equilibrium evolutions for different prescriptions. 

Summarizing, we have presented a compact path integral formalism to deal with
{\em Markovian multiplicative white-noise} systems,  independently of the
pres\-crip\-tion used to define the Wiener integral. In its supersymmetric
covariant form, it automatically encodes equilibrium properties. Therefore, it
is very appropriate to organize perturbative calculations, since
supersymmetry imposes non-perturbative constraints that should be verified at any
order of perturbation theory.   On the other hand, keeping in checked equilibrium
properties, we can safely go forward in the study of
out-of-equilibrium fluctuation relations described by this type of stochastic
processes.   
 
\section*{Acknowledgedments}
The Brazilian agencies, {\em Funda\c c\~ao de Amparo \`a Pesquisa do Rio
de Janeiro}, FAPERJ and {\em Conselho Nacional de Desenvolvimento Cient\'\i
fico e Tecnol\'ogico}, CNPq are acknowledged  for partial financial support.
Z.G.A. was partially supported by the Latin American Center of Physics, CLAF,
under the collaborative doctoral program ICTP-CLAF.

%%%%%%%%%%%%%%%%%%%%%%%%%%%%%%%%%%%%%%%%%%%%%%%%
\begin{appendix}

\section{}
\label{appendix}
In this appendix we explicitly show that  the action, written in the super-field
formalism, equation~(\ref{eq.SSUSY}), 
is in fact equivalent to the time reversal invariant  
action $\tilde S$, given by equation~(\ref{eq.Stilde}).

Let us define the scalar super-field 
\begin{equation}
\Phi(t,\theta,\bar\theta) = x(t)+ \bar \theta\eta(t)+ \bar\eta(t)\theta  +i \tilde\varphi(t) \bar\theta \theta.
\end{equation}
and the super-potentials  $V(\Phi)$  and $\Gamma(\Phi)$, that we expand in
powers of the temporal Grassman variables $\theta$ and $\bar\theta$ as, 
\begin{equation}
\Gamma(\Phi)=  \Gamma(x)+\Gamma'(x) \left\{\bar \theta\eta(t)+ \bar\eta(t)\theta\right\}  + 
\bar\theta \theta\left\{i \tilde\varphi(t)\Gamma'(x)- \Gamma''(x)\bar\eta\eta \right\},
\end{equation}
and 
\begin{equation}
 V(\Phi)=  V(x)+V'(x) \left\{\bar \theta\eta(t)+ \bar\eta(t)\theta\right\}  + 
\bar\theta \theta\left\{i \tilde\varphi(t)V'(x)- V''(x)\bar\eta\eta \right\}.
\label{eq.VPhi}
\end{equation}

The covariant derivatives (\ref{eq.cov_derivs}) acting on the super-potential
$\Gamma(\Phi)$ take the form 
\begin{equation}
\bar D\Gamma(\Phi)=  -\Gamma'(x)\bar\eta - 
\bar\theta\left\{i \tilde\varphi(t)\Gamma'(x)- \Gamma''(x)\bar\eta\eta \right\}
\label{eq.DGammabar}
\end{equation}
and
\begin{eqnarray}
D\Gamma(\Phi)&=& \frac{1}{2}\Gamma'(x)\eta + \frac{1}{2} 
\theta\left\{i \tilde\varphi(t)\Gamma'(x)- \Gamma''(x)\bar\eta\eta - 2\frac{d}{dt} \Gamma(x) \right\}+ 
\nonumber \\
&+&\bar\theta\theta\frac{d}{dt}\left(\Gamma'(x)\eta\right).
\label{eq.DGamma}
\end{eqnarray}

The system Lagrangian is given by the $\bar\theta\theta$ components in the
form
\begin{equation}
L=\left\{\bar D\Gamma D\Gamma+\frac{1}{2} V\right\}_{\bar\theta\theta}.
\label{eq.Lagrangian}
\end{equation}
Then, using equations (\ref{eq.VPhi}), (\ref{eq.DGammabar}) and (\ref{eq.DGamma}),
\begin{eqnarray}
\left.\bar D\Gamma D\Gamma\right|_{\bar\theta\theta}&=& -\bar\eta\Gamma'\frac{d}{dt}\left(\Gamma'\eta\right) +  
i \tilde\varphi \left\{ \Gamma'\frac{d}{dt}\Gamma+ \Gamma'\Gamma\bar\eta \eta\right\}-\nonumber \\
&-&\Gamma''\frac{d}{dt}\Gamma\bar\eta \eta +
\frac{1}{2} \tilde\varphi^2\Gamma'^2
\end{eqnarray}
and
\begin{equation}
\left.V(\Phi)\right|_{\bar\theta\theta}=i\tilde\varphi(t) V'(x)-V''(x)\bar\eta \eta.
\end{equation}
Replacing these expressions into equation (\ref{eq.Lagrangian}) we find
\begin{eqnarray}
L&=& -\bar\eta\Gamma'\frac{d}{dt}\left(\Gamma'\eta\right) + 
i \tilde\varphi \left\{ \Gamma'\frac{d}{dt}\Gamma+ \frac{1}{2}V'+ \Gamma'\Gamma''\bar\eta \eta\right\}-\nonumber \\
&-&\Gamma''\frac{d}{dt}\Gamma \bar\eta \eta+
\frac{1}{2}\tilde\varphi^2{\Gamma'}^2-\frac{1}{2} V''(x) \bar\eta \eta
\end{eqnarray}
We make the following change of variables in the functional integral:
first, we shift the variable $\tilde\varphi$, 
\begin{equation}
\tilde\varphi(t)\to \varphi_1(t) -i 2 \frac{\Gamma''(x)}{\Gamma'(x)}
\end{equation} 
whose Jacobian is one, since it is just a translation. 
Then, we rescale $(\varphi_1,\eta,\bar\eta)$ in the following way,  
\begin{eqnarray}
\eta(t) &\to&  (\Gamma'(x))^{-1}\xi(t)\\
\bar\eta(t) &\to&  \bar\xi(t) (\Gamma'(x))^{-1} \\
\varphi_1(t)&\to&\varphi(t) (\Gamma'(x))^{-2}\;.
\end{eqnarray}
The Jacobian of the first two transformations is $\det^2(\Gamma')$ and the Jacobian of the last transformation is 
$\det^{-2}(\Gamma')$, then the two Jacobians cancel each other and the Lagrangian in the  new variables reads, 
\begin{eqnarray}
L&=& - \bar\xi\dot \xi + \frac{1}{2}\frac{\varphi^2}{{\Gamma'}^2} + i \varphi \left\{ \frac{1}{{\Gamma'}}\frac{d}{dt}{\Gamma} + 
\frac{1}{2}\frac{V'}{{\Gamma'}^2}- \frac{{\Gamma''}}{{\Gamma'}^3}\bar\xi\xi\right\}-\nonumber \\
&-&\frac{1}{2}\left(\frac{V'(x)}{{\Gamma'}^2}\right)' \bar\xi\xi +\frac{{\Gamma''}}{{\Gamma'}^3}\frac{d}{dt}{\Gamma}\bar\xi\xi.
\label{eq.Lint}
\end{eqnarray}

Considering that, by definition,  ${\Gamma'}(x)=1/g(x)$ and, thus, ${\Gamma''}(x)=-g'(x)/g^2(x)$, we  obtain
\begin{eqnarray}
L&=& -\bar\xi\dot \xi + \frac{1}{2}g^2\varphi^2 +
i \varphi \left\{ \dot x+ \frac{1}{2}g^2 V'+ g g'\bar\xi\xi\right\}-\nonumber \\
&-&\frac{1}{2}\left(g^2 V'\right)' \bar\xi\xi -\frac{1}{2}\frac{d\ln g^2}{dt}\bar\xi\xi-\frac{1}{2}\frac{d}{dt} V.
\end{eqnarray}
The last term is a total derivative. This term appears making  a trivial transformation like $t\to t+1/2 \bar\theta\theta$ in the $V(\Phi)$ term.   
This is because a supersymetric Lagrangian is invariant up to a total derivative under SUSY transformations.
This is a general feature of SUSY. A transformed supersymmetric action is invariat up to boundary terms.

Rewriting this last expression in terms of the original functions $f$ and $g$ we finally get
\begin{equation}
L = - \bar\xi\dot \xi + \frac{1}{2}g^2\varphi^2 + i \varphi \left\{ \dot x- f+ g g'\bar\xi\xi\right\}
+ \bar\xi\xi f' -\frac{1}{2}\frac{d~}{dt}\left\{  V+\bar\xi\xi\ln g^2\right\}.
\end{equation}

Upon integration and realizing that the last term is a total derivative one, we find 
\begin{equation}
S_{SUSY}=\int_{t_i}^{t_f} dt\; L = S-\frac{1}{2}\left(V+\alpha\; \ln g^2\right)_{t_i}^{t_f}=\tilde S.
\label{eq.Lfinal}
\end{equation}
Where $S$ is the action of equation (\ref{eq.SGrassman}) and, then, $\tilde S$ is equal to equation (\ref{eq.Stilde}), as can be seen comparing 
(\ref{eq.Lfinal}) with (\ref{eq.SST}), as we want to demonstrate. 

To compute correlation functions the range of integration is infinite and $S=\tilde S$. However, for finite time we see that the supersymmetric action 
is equivalent to the time reversal invariant action $\tilde S$.

\end{appendix}

\section*{References}
%\bibliographystyle{iopart-num}
%\bibliography{stochastic}

\providecommand{\newblock}{}

\end{document}